\documentstyle[12pt]{article}
\title{\bf Dissipative particle dynamics with energy conservation:
equilibrium properties}
\author{Allan D. Mackie and Josep Bonet Avalos\\
Departament d'Enginyeria Qu\'{\i}mica, ETSEQ \\
Universitat Rovira i Virgili.\\ Carretera de Salou s/n, 43006
Tarragona (Spain)}

\date{}

\begin{document}
\maketitle
\parskip 2ex
\renewcommand{\theequation}{\arabic{section}.\arabic{equation}}

\renewcommand{\baselinestretch}{1.5}

\baselineskip=1.5\baselineskip

\hrule

\begin{abstract} 
The stochastic differential equations for a model of
dissipative particle dynamics, with both total energy and total momentum
conservation at every time-step, are presented. The algorithm satisfies
{\em detailed balance} as well as the {\em fluctuation-dissipation}
theorems that ensure that the proper thermodynamic equilibrium can be
reached. Macroscopic equilibrium probability distributions as well as {\em
equations of state} for the model are also derived, and an appropriate 
definition of the free energy of the system is consistently proposed. Several
simulations results of equilibrium as well as transport properties,
including heat transport and thermal convection in a box, are shown as
proof of the internal consistency of the model.
\end{abstract}

\hrule

\newpage

\section{Introduction}

\setcounter{equation}{0}        

        The computer simulation
strategy for dynamics of complex systems known as {\em
Dissipative Particle Dynamics} or, simply, DPD, has been the subject of
several studies in recent years. This methodology was introduced by 
Hoogerbrugge {\em et al.}\cite{Hoo} to model the dynamic behaviour of
fluids by using a particulate method in which an ensemble of {\em mesoscopic}
particles interact with each other via conservative as well as dissipative
forces. 
Contrary to other particulate {\em Lagrangian}
methods\cite{Mon}, oriented at modelling the hydrodynamics of macroscopic
systems, 
DPD also incorporates Brownian forces in the particle-particle
interactions. In this way, thermal fluctuations can be described and a certain
thermodynamic equilibrium exists. DPD is, therefore, especially
suited to the modelling of mesoscopic 
systems, where fluctuations play an important role. One can mention, for
instance, the dynamics of polymer solutions, nucleation and phase separation
problems, dynamics of confined fluids, colloidal suspensions, and
kinetics of chemical reactions in 
confined geometries, etc. The DPD method has already been succesfully
applied to 
some cases of practical interest such as to the rheology of
colloidal suspensions\cite{Lek} and to the microphase separation of diblock
copolymer solutions\cite{Mad}, among others.

        In DPD, dissipative and
random interactions are pairwise and chosen in such a way that the center
of mass 
motion of each interacting pair is insensitive to these frictional and
random forces. Hence, on the one hand, if closed and thermally isolated,
the system relaxes fast to its thermal equilibrium and, on the 
other, its overall momentum is a conserved variable in view of the
particle's momentum exchange rules. This second feature 
allows the system to exhibit a hydrodynamic behaviour from
a macroscopic point of view, that is, for length and time scales larger
than those characterizing the particle-particle interactions. With respect
to other mesoscopic models for hydrodynamic behaviour such as Lattice
Gas\cite{who} or Cellular 
Automata\cite{whoelse}, the DPD model is isotropic and
Galilean-invariant due to the fact that it is not defined on a lattice and
is straightforwardly based on Newton's equations of motion, as in MD. In
addition, it has no extra conservation laws and is also computationally
efficient. As originally formulated, however, the 
DPD model can only deal with isothermal conditions since dissipative and 
Brownian forces cause no energy conservation in the particle-particle
interaction. The equation for the energy transport is then of
the relaxation type\cite{Ern} and, therefore, heat flow, related to the
energy conservation at the microscopic scale, cannot be described within
the framework of the original model (which will be referred to as
isothermal DPD from now on). In many problems of interest, either
fundamental or applied, the study of the transport properties of heat at the
mesoscopic level are very important. Thus, the incorporation of the
thermal effects into a DPD algorithm is necessary for future
applications of the method to problems of relevance. 

        In a previous
paper\cite{ours} we introduced an algorithm in which the conservation of
the total
energy in the particle-particle interaction was consistently taken into
account with the inclusion of the particle's internal energy
in the model (this model will be referred to as DPDE from now on). In this
paper we will analyze several aspects of this novel DPDE algorithm. In the
first place, we discuss the 
extension of our original algorithm, given in ref.\cite{ours}, to
incorporate, on the one hand,
temperature dependencies in the dynamic parameters of the model so that the
treatment of  arbitrary temperature-dependencies in the transport
coefficients can be treated, something lacking in the older DPD models.
On the other hand, we develop a direct derivation of the algorithm from
the Langevin equations of 
motion for the relavant variables, by introducing a interpretation rule
for these Langevin equations that ensures that the detailed balance
condition is satisfied in an Euler-like algorithm. The detailed balance
condition is necessary for the system to evolve
towards the proper thermodynamic equilibrium. Furthermore, 
the algorithm obtained in this way conserves the
energy at {\em every time-step} instead of {\em in the mean} as was the
case in the older DPDE 
algorithm\cite{ours}. We have noticed that different
algorithms lead to the same Fokker-Planck equation, i.e., different
stochastic processes can lead, in fact, to the same dynamics for the
probability 
distribution. In the second place, we have analyzed the macroscopic
equilibrium properties of the DPDE system. We have shown that a
thermodynamic analysis of the DPDE system is possible in view of the fact
that the equilibrium probability distribution is known. Properties
such as a free energy or the equations of state can be derived in terms of
the parameters defining the model. The simulation results of the
equilibrium properties show an excellent agreement with the theoretical
predictions. Third and last, we have done simulations in systems under a
temperature gradient imposed by the temperature of two walls. We have
found temperature profiles, as expected since the model allows the
simulation of heat transport, and the existence of inhomogeneous
temperature distributions, which is one of the main virtues of the DPDE
algorithm. Furthermore, if a gravity field is considered, the system
undergoes convective motion in addition to the temperature gradient, the
Rayleigh number qualitatively estimated as being of the order of $10^3$.

        The paper is organized as follows. In section II, we introduce the
main hypothesis underlying the 
formulation of the DPDE model\cite{ours}, and derive the
appropriate algorithm with arbitrary dependence of the transport
coefficients with the temperature, and energy conservation at every
time-step. The  
Fokker-Planck equation describing the evolution of the probability
distribution for the ensemble of variables describing the state of the
system is also derived together with the corresponding
fluctuation-dissipation theorems. In 
section III, we obtain the macroscopic equilibrium properties of the DPDE
model and 
expressions for the equations of state and thermodynamic
properties. At the end of this section, we pay
attention to a particular model, used to perform simulations of
equilibrium properties as well as heat
transport and thermal convection. Finally, in
section IV we draw the main conclusions from our work. 

\section{Dissipative particles with energy conservation}

        For our purposes, it is useful to have a physical picture and
regard the dissipative particles as if they were {\em clusters}
of true physical particles\cite{Esp1,Ern}, i.e., as particles with 
internal structure bearing some degrees of freedom. Thus, the DPDE model is
mesoscopic in nature since it resolves only the overall center-of-mass
motion of the cluster and 
ignores the exact {\em internal state} of the cluster as a relevant variable.
The interactions being dissipative and random, however, the total energy
of the system is not conserved unless the energy exchanged between the
resolved degrees of freedom and the internal
state of the DPD particle is accounted for. We propose here a model based
on the treatment of 
thermodynamic and hydrodynamic fluctuations\cite{Ca,dGr,vSa}, to
consistently take into account the energy stored in the internal degrees
of freedom of each particle, without explicit consideration of any {\em
internal} Hamiltonian. 

\subsection{Definition of the model}

        Our model is based on the following assumptions:  
\begin{enumerate}
\item The system contains $N$ particles interacting with each other via
conservative as well as dissipative interactions. The conservative
interactions are described by the Hamiltonian
\begin{equation}
H(\{\vec{r}_i\},\{\vec{p}_i\}) \equiv \sum_i^N \left\{ \frac{p_i^2}{2m}
+\sum_{j>i} \psi (r_{ij}) +\psi^{ext}(\vec{r}_i) \right\}          \label{1}
\end{equation}
where $r_{ij} \equiv |\vec{r}_i-\vec{r}_j|$. The Hamiltonian depends on
the momenta and the positions of all the particles. 
The particles interact  through
pair pontentials $\psi (|\vec{r}_i-\vec{r}_j|)$, depending only on the
distance between them, and with an external field
$\psi^{ext}(\vec{r}_i)$ 
\item In addition, the particles can store energy in some internal degrees
of freedom. The internal energy $u_i$, with $u_i \geq 0$, is introduced as a
new relevant coordinate. The momentum $\vec{p}_i$, the
position $\vec{r}_i$ together with the internal energy $u_i$, completely
specify the state of the dissipative particle at a given instant $t$.
\item The particle-particle interaction is pairwise and conserves the
total momentum and the total energy when the internal energy of the pair
is taken into account. 
\item The internal states of the particle have no dynamics in the sense
that they are always in equilibrium with themselves. This allows us to
define a function $s_i(u_i)$. This
function can arbitrarily be chosen according to the user's needs, except
that it is
constrained to thermodynamic consistency requirements\cite{Ca}
\begin{enumerate}
\item $s_i$ must be a differentiable monotonically increasing function of
its variable $u_i$, so that $u_i(s_i)$ exists and $\theta_i \equiv
\partial u_i/\partial s_i$ exists and is always positive.
\item $s_i$ is a concave function of its argument.
\end{enumerate}
Defined in this way, $s_i$ can be viewed as a mesoscopic {\em entropy} of
the $i^{th}$ particle, and $\theta_i$ can be seen as the 
particle's {\em temperature}. The change in
$u_i$ and in $s_i$ are related by a Gibbs equation 
\begin{equation}
\theta_i \, ds_i =
du_i,\;\; \mbox{which implies} \;\;  \theta_i \, \dot{s_i} = \dot{u_i}
                        \label{Gibbs} 
\end{equation}
where the dot over the variables is used to denote time-differentiation 
from now on.
\item The irreversible particle-particle
interaction is such that the deterministic part (in the
absence of random forces) must satisfy
\begin{equation}
\dot{s_i} + \dot{s_j} \geq 0            \label{secondlaw}
\end{equation}
where $i$ and $j$ label an arbitrary pair of interacting particles.
\item $u_i$, $s_i$ and $\theta_i$ must
remain unchanged under a Galilean transformation, so that these variables
are true scalars.
\item The equilibrium probability distribution
for the relevant variables of the system under isothermal conditions is
chosen to be 
\begin{equation}
P_e(\{\vec{r}_i\},\{\vec{p}_i\},\{u_i\}) \sim
e^{-H(\{\vec{r}_i\},\{\vec{p}_i\})/kT}\prod_i e^{s_i(u_i)/k-u_i/kT}
                \label{Pe}
\end{equation}
where $k$ is Boltzmann's constant and $T$ is the thermodynamic, i.e.,
macroscopic 
temperature. The first factor on the right hand side of eq. (\ref{Pe})
contains the probability distribution for the set of variables
$\{\vec{r}_i\},\{\vec{p}_i\}$, as given by equilibrium statistical
mechanics. The second factor on the right hand side corresponds, in turn,
to the 
probability distribution for the internal energy of the particles as
obtained from equilibrium fluctuation theory\cite{Ca}. Effectively,
$\exp(s_i(u_i)/k-u_i/kT)$ gives the probability for the $i^{th}$ particle
to have an internal energy $u_i$ regarding the rest of the system as a heat
reservoir. The maximum of $s_i(u_i)/k-u_i/kT$ occurs at $\theta_i =
T$, in agreement with our interpretation of $\theta_i$ as the particle's
temperature. Furthermore, notice that
\begin{equation}
\left \langle \frac{1}{\theta_i} \right \rangle = \frac{1}{\cal N} \int du_i \;
e^{(s_i(u_i)/k-u_i/kT)} \frac{\partial s_i}{\partial u_i} = \frac{1}{T} 
\end{equation}
where ${\cal N}$ is the normalisation constant.
Once the equilibrium probability distribution is obtained, the {\em
thermodynamic} properties of the model are determined. 

\end{enumerate}

\subsection{Dynamics of the model}

        The model defined so far must be completed with a set of
equations that explicitly establish its dynamical properties. Initially, in
view of the pairwise additivity of the interactions, we will analyze an
arbitrary  
pair of particles, $i$ and $j$ say, and later on give the complete
expressions for the $N$-particle system.

        The change in position and momentum of the
$i^{th}$ particle due to the interaction with the $j^{th}$ particle
follows from Newton's second law
\begin{eqnarray}
\dot{\vec{r}}_i &=& \frac{\vec{p}_i}{m}   \label{4}\\
\dot{\vec{p}}_i &=& \vec{F}_{ij}^C+ \vec{F}_i^{ext} + \vec{F}_{ij}^D
+\vec{F}_{ij}^R   \label{5}
\end{eqnarray}
where $\vec{F}_{ij}^C \equiv -\partial \psi(r_{ij})/\partial \vec{r}_i$ and
$\vec{F}_i^{ext} \equiv -\partial \psi^{ext}(\vec{r}_i)/\partial \vec{r}_i$ are the
forces due to the conservative interactions. $\vec{F}_{ij}^C$ is, by
construction, directed along the vector $\hat{r}_{ij} \equiv
(\vec{r}_j-\vec{r}_i)/r_{ij}$ and satisfies
$\vec{F}_{ij}^C=-\vec{F}_{ji}^C$. In addition, $\vec{F}_{ij}^D$ stands for the 
dissipative particle-particle interaction force and $\vec{F}_{ij}^R$ is
the random force associated with the former. The total energy $e_i$ of the
$i^{th}$ particle is the sum of the kinetic ($p_i^2/2m$), the
potential ($\psi(r_{ij})/2+\psi^{ext}(\vec{r}_i)$), and the internal energy
($u_i$) contributions. Using eq. 
(\ref{5}) to compute the change in the kinetic energy, we arrive at the
equation for the change in the
total energy
\begin{equation}
\dot{e}_i = \frac{d}{dt} \frac{p_i^2}{2m} + \frac{d}{dt}
\left(\frac{1}{2}\psi(r_{ij})+\psi^{ext}(\vec{r}_i) \right) + \dot{u}_i
=\frac{1}{m} 
\vec{p}_i \cdot \left(\vec{F}_{ij}^D+ \vec{F}_{ij}^R \right)+ \frac{1}{2m} 
(\vec{p}_i+\vec{p}_j) \cdot \vec{F}_{ij}^C + \dot{u}_i.
        \label{6} 
\end{equation}

        The conservation of the total momentum of the pair imposes that
its change is only due to the action of external force fields, that is,
\begin{equation}
\dot{\vec{p}}_i+\dot{\vec{p}}_j = \vec{F}_i^{ext}+ \vec{F}_j^{ext},
        \label{7} 
\end{equation}
Therefore, we must have that $\vec{F}_{ij}^{D,R} = -\vec{F}_{ji}^{D,R}$.
In addition, 
conservation of the total angular momentum
\begin{equation}
\vec{r}_i \times \dot{\vec{p}}_i+\vec{r}_j \times \dot{\vec{p}}_j =
\vec{r}_i \times \vec{F}_i^{ext}+ \vec{r}_j \times \vec{F}_j^{ext}
        \label{8} 
\end{equation}
implies that dissipative as well as Brownian forces, $\vec{F}_{ij}^{D,R}$,
must be directed along the unit vector 
$\hat{r}_{ij}$. Conservation of the total energy in the particle-particle
interaction, $\dot{e}_i+\dot{e}_j = 0$ gives, in turn, the rate of change
of the internal energy of the pair
\begin{equation}
\dot{u}_i+\dot{u}_j = -\frac{1}{m} ( \vec{p}_i - \vec{p}_j) \cdot
\left( \vec{F}_{ij}^D+\vec{F}_{ij}^R \right)          \label{9}
\end{equation}
This equation implies that if the total energy is to be conserved, then the
change in the total internal energy has to be due to the work done by the
dissipative and the associated random
forces. The dissipative forces transfer mechanical energy to internal
energy, while random forces bring back internal energy to the kinetic and
potential energies of the particles.
From the conservation equations alone, however, nothing can be inferred
about how the internal energy is distributed among the particles so that
one has to supply a given model. 

        We will assume that the
mechanisms driving the change in the internal energy of the particles are
of two kinds. On the one hand, the work done by dissipative and random
forces is shared in 
equal amounts by the particles and is irrespective of their
temperatures $\theta_i$ and $\theta_j$. On the other hand, we assume that
the particles can also vary 
their internal energy
by exchanging internal energy if $\theta_i \neq \theta_j$. The energy
tranferred 
by this mechanism between the particles will be referred to as {\em
mesoscopic heat flow} $\dot{q}_{ij}^D$. Associated with this dissipative
current, a random heat flow $\dot{q}_{ij}^R$ must also be added. Hence, we
write
\begin{equation}
\dot{u}_i = -\frac{1}{2m} ( \vec{p}_i - \vec{p}_j) \cdot
\left(\vec{F}_{ij}^D+\vec{F}_{ij}^R \right)  + \dot{q}_{ij}^D +
\dot{q}_{ij}^R  \label{10}
\end{equation}
with the requirement $\dot{q}_{ij}^{D,R} = -\dot{q}_{ji}^{D,R}$ to ensure
that eq. (\ref{9}) is satisfied. Note that the r.h.s. of eq. (\ref{10})
preserves Galilean-invariance. 

        So far, only the conservation equations have been used to find
general properties to be satisfied by the dissipative forces and
mesoscopic heat flows, referred to as {\em dissipative currents} in a wider
sense from now on. In analogy with
the Thermodynamics of Irreversible Processes\cite{dGr}, we can make use of
the Gibbs equation, eq. (\ref{Gibbs}) together with eq. (\ref{10}) to find
the {\em particle's entropy production} from the interaction between
pairs, yielding 
\begin{equation}
\dot{s}_i+\dot{s}_j =
\frac{\dot{u}_i}{\theta_i}+\frac{\dot{u}_j}{\theta_j} =
-\frac{1}{m}\left(\frac{1}{\theta_i}+\frac{1}{\theta_j} \right)
(\vec{p}_i-\vec{p}_j) \cdot 
\vec{F}_{ij}^D+\left(\frac{1}{\theta_i}-\frac{1}{\theta_j} \right) 
\dot{q}_{ij}^D \geq 0     \label{11}
\end{equation}
where only the deterministic part of the interactions has explicitly been
considered, according to point 5. The inequality in this last equation is
the expression of the  
Second Law of Thermodynamics, indicating the irreversibility of the
particle's dissipative interations.
From the entropy production equation the so-called {\em thermodynamic
forces} can be identified as the factors multiplying the respective
dissipative currents, $\vec{F}_{ij}^D$ and $\dot{q}_{ij}^D$ in eq.
(\ref{11}). Thus, in the spirit of the
Thermodynamics of Irreversible Processes\cite{dGr} we propose a linear
relation 
between the dissipative currents and the thermodynamic forces of the form
\begin{eqnarray}
\vec{F}_{ij}^D &=& L_{ij}^{(p)}
\frac{1}{m}\left(\frac{1}{\theta_i}+\frac{1}{\theta_j} \right)
\hat{r}_{ij}\hat{r}_{ij} \cdot (\vec{p}_j-\vec{p}_i)  \label{12} \\
\dot{q}_{ij} &=& L_{ij}^{(q)} \left(\frac{1}{\theta_i}-\frac{1}{\theta_j} \right)
                \label{13} 
\end{eqnarray}
The functions $L_{ij}^{(p)}$ and $L_{ij}^{(q)}$ are analogous to the
so-called Onsager coefficients\cite{dGr}. Due to the different
tensorial natures of the momentum flux and the heat flux, these phenomena
are not coupled in eqs. (\ref{12}) and (\ref{13}) 
(Curie's theorem\cite{dGr}). The mesoscopic Onsager
coefficients introduced in eqs. (\ref{12}) and (\ref{13}) must satisfy the
following conditions so that the system can reach the proper thermodynamic
equilibrium
\begin{enumerate}
\item The thermodynamic forces are Galilean-invariant, as are the
dissipative currents. This implies that the Onsager coefficients must also
be Galilean-invariant.
\item Microscopic reversibility implies that $L_{ij}^{(p)}$ and
$L_{ij}^{(q)}$ must be even functions under time-reversal\cite{JBA}.
\item Since the thermodynamic forces change their sign under the exchange $i
\rightarrow j$, then $L_{ij}^{(p)}$ and $L_{ij}^{(q)}$ must be {\em invariant}
under this transformation. This fact is crucial in the derivation of the
transport properties of the system.
\end{enumerate}

        In the {\em linear} Non-Equilibrium Thermodynamics scheme,
$L_{ij}^{(p)}$ and $L_{ij}^{(q)}$ are constants. However, this
choice restricts the expression for the macroscopic transport
coefficients of the model to a given 
particular form which, in addition, seldom occurs in nature\cite{vSa}. Thus,
we will consider a general dependence of the Onsager coefficients in both
the distance $r_{ij}$ and the temperatures of the particles $\theta_i$ and
$\theta_j$, in order to allow the model to describe the temperature
dependence of the macroscopic transport coefficients. Thus, for
convenience, let us rewrite eqs. (\ref{12}) and 
(\ref{13}) in a different form
\begin{eqnarray}
\vec{F}_{ij}^D &=& \zeta_{ij} \frac{1}{m}
\hat{r}_{ij}\hat{r}_{ij} \cdot (\vec{p}_j-\vec{p}_i)  \label{14} \\
\dot{q}_{ij} &=& \lambda_{ij} (\theta_j-\theta_i)
                \label{15} 
\end{eqnarray}
Written in this way, these expressions are reminiscent of the macroscopic
phenomenological Newton's law for the transport of momentum, and
Fourier's law for heat transport\cite{dGr}. Hence, $\zeta_{ij}$ is a
mesoscopic analog of the macroscopic
viscosity and $\lambda_{ij}$, of the thermal conductivity. $\zeta_{ij}$ and
$\lambda_{ij}$ will be referred to as {\em mesoscopic} dissipative
coefficients, from now on. Clearly, the mesoscopic dissipative coefficients
are related to 
the mesoscopic Onsager coefficients introduced in eqs. (\ref{12}) and
(\ref{13}), according to
\begin{eqnarray}
L_{ij}^{(p)} &=& \Theta_{ij} \zeta_{ij} \label{16} \\
L_{ij}^{(q)} &=& \theta_i \theta_j \, \lambda_{ij}    \label{17}
\end{eqnarray}
where $\Theta_{ij}^{-1} \equiv (1/\theta_i+1/\theta_j)/2$. In what
follows, we will use the mesoscopic dissipative coefficients
instead of the mesoscopic Onsager coefficients, to base our discussion in
the most intuitive manner. From eqs. (\ref{16}) and
(\ref{17}) one can see that if
$\zeta_{ij}$ and $\lambda_{ij}$ are constants, then the mesoscopic Onsager
coefficients depend on the particles' internal energy through the
temperatures $\theta_i$ and $\theta_j$. The presence of
temperature-dependencies in the mesoscopic Onsager coefficients will introduce
subtleties in the derivation of the algorithm due to the so-called
It\^o-Stratonovich dilemma\cite{vK} that we will discuss later on.

        The properties of the random terms are also chosen to parallel the
theory of hydrodynamic fluctuations\cite{vSa,Lan}. Since $\vec{F}_{ij}^D$ and
$\dot{q}_{ij}^D$ are not coupled, we will demand that the random terms
$\vec{F}_{ij}^R$ and $\dot{q}_{ij}^R$ be statistically independent. They
can be written in the form
\begin{equation}
\vec{F}_{ij}^R = \hat{r}_{ij} \, \Gamma_{ij} \, {\cal F}_{ij}(t)  \;\;\;
\mbox{and} \;\;\; \dot{q}_{ij}^R = \mbox{Sign}(i-j) \, \Lambda_{ij} \, {\cal
Q}_{ij}(t)         \label{18}
\end{equation}
where the function Sign$(i-j)$ is $1$ if $i>j$ and $-1$ if $i<j$, ensuring
that $\dot{q}_{ji}^R = -\dot{q}_{ij}^R$.
The scalar random variables ${\cal F}_{ij}$ and ${\cal Q}_{ij}$ are
stationary, Gaussian and white\cite{JBA,Maz2}, with zero mean and
correlations 
\begin{eqnarray}
\langle {\cal F}_{ij}(t) {\cal F}_{kl}(t') \rangle &=& \langle {\cal Q}_{ij}
(t) {\cal Q}_{kl} (t') \rangle = (\delta_{ik}\delta_{jl}
+\delta_{il}\delta_{jk} ) \, \delta (t-t')     \label{19} \\
\langle {\cal F}_{ij}(t) {\cal Q}_{kl}(t') \rangle &=& \langle {\cal Q}_{ij}
(t) {\cal F}_{kl} (t') \rangle = 0 \label{19b}
\end{eqnarray}
$\Gamma_{ij}$ and $\Lambda_{ij}$ are functions to be determined later.
Note that $\zeta_{ij}$ and $\Gamma_{ij}$ are, respectively, $\gamma
\omega_D$ and $\sigma \omega_R$ in ref.\cite{Esp1}.

\subsubsection{Derivation of the algorithm}

        As we have seen, the dynamics of the model is described by the
set of Langevin equations, eqs. (\ref{4}), (\ref{5}) and (\ref{10}), the
definition of the dissipative currents, eqs.
(\ref{14}) and (\ref{15}), and the random forces, eqs. (\ref{18}).
To derive an algorithm describing such a stochastic process, let us
integrate these Langevin equations for a small increment of time $\delta
t$ and retain terms up to ${\cal O}(\delta t)$. One obtains
\begin{eqnarray} 
\vec{r}_i^{\, \prime} &=& \vec{r}_i + \frac{\vec{p}_i}{m} \delta t 
                \label{21a} \\
\vec{p}_i^{\, \prime}  &=& \vec{p}_i +\left\{\vec{F}_{i}^{ext} + 
\sum_{j \neq i} \left[\vec{F}_{ij}^C + \frac{\zeta_{ij}}{m} \,
(\vec{p}_j-\vec{p}_i) \cdot \hat{r}_{ij} \hat{r}_{ij} \right] \right\}
\delta t +  
\sum_{j \neq i} \hat{r}_{ij} \Gamma_{ij}^{\prime} \, \delta
t^{1/2} \; 
\Omega_{ij}^{(p)} \nonumber \\
	\label{21b} \\  
u_i^{\prime} &=& u_i + \sum_{j \neq i} \left\{\frac{\zeta_{ij}}{2m^2}
\left[(\vec{p}_j-\vec{p}_i) \cdot \hat{r}_{ij} \right]^2 + \lambda_{ij}
\left(\theta_j-\theta_i \right)
\right\} \delta t  \nonumber \\ 
&+& \sum_{j \neq i} \left\{ \frac{1}{2m} 
(\vec{p}_j^{\, \prime}-\vec{p}_i^{\, \prime}) \cdot 
\hat{r}_{ij} \Gamma_{ij}^{\prime} \, 
\delta t^{1/2} \; \Omega_{ij}^{(p)} + \mbox{Sign}(i-j) \Lambda_{ij}^{\prime} \,
\delta t^{1/2} \; \Omega_{ij}^{(q)} \right\}      \label{21c}
\end{eqnarray}
where $\vec{r}_i^{~\prime} \equiv \vec{r}_i (t+\delta t)$,
$\vec{p}_i^{~\prime} 
\equiv \vec{p}_i (t+\delta t)$, and $u_i^{\prime} \equiv u_i (t + \delta t)$
while $\vec{r}_i$, $\vec{p}_i$ and $u_i$ are the value of these functions
at time $t$, and $\Gamma_{ij}^{\prime}$ and
$\Lambda_{ij}^{\prime}$ stand for the value of these functions when their
arguments are calculated at the time $t+ \delta t$.
The integrals over the random terms have been interpreted as\cite{vK}
\begin{equation}
\int_t^{t+ \delta t} d\tau \; \hat{r}_{ij}(\tau) \Gamma_{ij}[\tau] {\cal
F}_{ij}(\tau) = \hat{r}_{ij}(t) \Gamma_{ij}[t+\delta t] \, \delta
t^{1/2} \, \Omega_{ij}^{(p)}       \label{21d}
\end{equation}
in eq. (\ref{21b}), and 
\begin{eqnarray}
  &\int_t^{t+ \delta t} d\tau&  (\vec{p}_j(\tau)-\vec{p}_i(\tau)) \cdot 
\hat{r}_{ij}(\tau) \, \Gamma_{ij}[\tau] \, {\cal F}_{ij}(\tau) \nonumber \\ 
 & = & (\vec{p}_j(t+\delta t)-\vec{p}_i(t+\delta t)) \cdot \hat{r}_{ij}(t) \,
\Gamma_{ij}[t+\delta t] \, \delta t^{1/2} \, \Omega_{ij}^{(p)} 
                 \label{21e} \\
 &  \int_t^{t+ \delta t} d\tau & \mbox{Sign}(i-j) \Lambda_{ij}[\tau] \,
{\cal Q}_{ij}(\tau) = \mbox{Sign}(i-j)
\Lambda_{ij}[t+\delta t] \,  \delta t^{1/2} \; \Omega_{ij}^{(q)} \label{21f}
\end{eqnarray}
respectively, in eq. (\ref{21c}). In these equations, we have defined the
random numbers 
\begin{equation}
\Omega_{ij}^{(p)} \equiv \frac{1}{\delta t^{1/2}} \int_t^{t+ \delta t}
d\tau \; {\cal F}_{ij}(\tau) \;\; \mbox{and} \;\; \Omega_{ij}^{(q)} \equiv
\frac{1}{\delta t^{1/2}} \int_t^{t+ \delta t} d\tau \; {\cal Q}_{ij}(\tau)
        \label{21g}
\end{equation}
which are Gaussian, with zero mean and correlations $\langle
\Omega_{ij}^{(p)} \Omega_{kl}^{(p)} \rangle = \langle
\Omega_{ij}^{(q)} \Omega_{kl}^{(q)} \rangle = (\delta_{ik}\delta_{jl} +
\delta_{il}\delta_{jk})$ and $\langle
\Omega_{ij}^{(p)} \Omega_{kl}^{(q)} \rangle =0 $, according to eqs.
(\ref{19}) and (\ref{19b}).
Eqs. (\ref{21d}), (\ref{21e}) and (\ref{21f}) express the fact that the
amplitudes of the random forces may depend on the fast variables
themselves ($u_i$ through the temperature $\theta_i$, in this case),
introducing the so-called It\^o-Stratonovich dilemma\cite{vK}.
Here, we have
chosen an interpretation rule that is neither that of It\^o nor
Stratonovich, but 
has the property that the resulting stochastic process, once the
amplitudes of the random terms have been fixed, satisfies detailed
balance up to ${\cal O} (\delta t)$\cite{JBAN}. 
Note that in eqs. (\ref{21d}),
(\ref{21e}) and (\ref{21f}) the 
slow variable $\vec{r}_{ij}$ is taken at $t$ instead of at
$t+\delta t$, since the difference between both considerations is of order
${\cal O} (\delta t^{3/2})$; only fast variables, whose increments contain
terms of the order ${\cal O} (\delta t^{1/2})$, need to be taken at the
latest time since they introduce terms of order ${\cal O} (\delta t)$ in
the algorithm.
That the stochastic process introduced by eqs.(\ref{21a}), (\ref{21b}),
and (\ref{21c})
satisfies detailed balance can be verified from the functional form of the
resulting Fokker-Planck equation, since this is a property of the transition
probabilities and the equilibrium distribution\cite{vK}.
Other algorithms have been proposed that explicitly
incorporate time-reversibility in the equations of motion for the particles
\cite{Pag}, which eventually leads to the resulting stochastic process
satisfying detailed balance\cite{DB}. 
The forward
interpretation rule, however, permits us to straightforwardly relate the
differential equations for the change in the particle's variables
(Langevin equations) with an
Euler algorithm that will lead to the {\em proper thermodynamic
equilibrium} for the system.

        To obtain a causal or explicit form of the algorithm defined in
eqs. (\ref{21a}), (\ref{21b}) and (\ref{21c}), we expand the right hand
side of these equations in powers of $\delta t$ and retain terms of up to
first order. We then obtain
\begin{eqnarray}
\delta \vec{r}_i &=& \frac{\vec{p}_i}{m} \delta t       \label{A1} \\
\delta\vec{p}_i &=& \left\{\vec{F}_{i}^{ext} + 
\sum_{j \neq i} \left[\vec{F}_{ij}^C + \left(\frac{\zeta_{ij}}{m} +
\frac{1}{2m} \Gamma_{ij} \left(\frac{\partial}{\partial u_i} +
\frac{\partial}{\partial u_j} \right) \Gamma_{ij}
\Omega_{ij}^{(p)^{2}}\right)\, 
(\vec{p}_j-\vec{p}_i) \cdot \hat{r}_{ij} \hat{r}_{ij} \right. \right.
\nonumber \\ 
 & + & \left. \left. \hat{r}_{ij} \, 
\mbox{Sign} (i-j) \, \Lambda_{ij} \left(\frac{\partial}{\partial u_i} -
\frac{\partial}{\partial u_j} \right) \Gamma_{ij} \Omega_{ij}^{(p)}
\Omega_{ij}^{(q)} \right]
\right\} \delta t +  
\sum_{j \neq i} \hat{r}_{ij} \Gamma_{ij} \, \delta
t^{1/2} \; 
\Omega_{ij}^{(p)} \label{A2} \\  
\delta u_i &=& \sum_{j \neq i} \left\{\frac{1}{2m}
\left(\frac{\zeta_{ij}}{m} + 
\frac{1}{2m} \Gamma_{ij} \left(\frac{\partial}{\partial u_i} +
\frac{\partial}{\partial u_j} \right) \Gamma_{ij}
\Omega_{ij}^{(p)^{2}}\right) 
\left((\vec{p}_j-\vec{p}_i) \cdot \hat{r}_{ij} \right)^2 + \lambda_{ij}
(\theta_j-\theta_i) \right. \nonumber \\
 &+& \left. \frac{1}{2m} (\vec{p}_j-\vec{p}_i) \cdot
\hat{r}_{ij} \Omega_{ij}^{(p)} \Omega_{ij}^{(q)} \mbox{Sign} (i-j) \, 
\left[\Gamma_{ij} \left(\frac{\partial}{\partial u_i} +
\frac{\partial}{\partial u_j} \right) \Lambda_{ij} -\Lambda_{ij}
\left(\frac{\partial}{\partial u_j} - 
\frac{\partial}{\partial u_i} \right) \Gamma_{ij} \right] \right. \nonumber \\
 &-& \left.  \Lambda_{ij}
\left(\frac{\partial}{\partial u_j} - 
\frac{\partial}{\partial u_i} \right) \Lambda_{ij} \Omega_{ij}^{(q)^{2}}
-\frac{1}{m} \Gamma_{ij}^{2} \Omega_{ij}^{(p)^{2}} 
\right\} \delta t  \nonumber \\ 
&+& \sum_{j \neq i} \left\{ \frac{1}{2m} 
(\vec{p}_j-\vec{p}_i) \cdot 
\hat{r}_{ij} \Gamma_{ij} \, 
\delta t^{1/2} \; \Omega_{ij}^{(p)} + \Lambda_{ij} \,
\delta t^{1/2} \; \Omega_{ij}^{(q)} \right\}      \label{A3}
\end{eqnarray}
Eqs. (\ref{A1}), (\ref{A2}) and (\ref{A3}) is an explicit algorithm,
equivalent to the implicit algorithm given in eqs. (\ref{21a}),
(\ref{21b}) and 
(\ref{21c}) up to order ${\cal O} (\delta t)$. By using standard
methods\cite{vK}, from eqs. (\ref{A1}), (\ref{A2}) and (\ref{A3}) one
obtains the Fokker-Planck equation 
\begin{equation}
\frac{\partial}{\partial t} P(\{\vec{r}_i\},\{\vec{p}_i\},\{u_i\})= L^{con}
P(\{\vec{r}_i\},\{\vec{p}_i\},\{u_i\}) + L^{dif}
P(\{\vec{r}_i\},\{\vec{p}_i\},\{u_i\})          \label{19e}
\end{equation}
where the {\em convective} operator, $L^{con}$, and the {\em diffusive}
operator, $L^{dif}$, are defined as
\begin{eqnarray}
L^{con} & \equiv & -\sum_{i=1}^N \left[\frac{\partial}{\partial \vec{r}_i} \cdot
\frac{\vec{p}_i}{m} +\frac{\partial}{\partial \vec{p}_i} \cdot
\vec{F}_i^{ext} \right]-\sum_{i,j\neq i}^N \left\{ \frac{\partial}{\partial
\vec{p}_i} \cdot \left[ \vec{F}_{ij}^C + \frac{\zeta_{ij}}{m}
\hat{r}_{ij}\hat{r}_{ij} \cdot (\vec{p}_j-\vec{p}_i) \right] + \right. 
\nonumber \\
& + & \left. \frac{\partial}{\partial u_i} \left[\frac{\zeta_{ij}}{2m^2}
\left[(\vec{p}_j-\vec{p}_i)\cdot \hat{r}_{ij} \right]^2 + \lambda_{ij}
\left(\theta_j -\theta_i \right) \right] \right\}
\label{19f} \\
L^{dif} & \equiv & \sum_{i,j\neq i}^N \left\{ \frac{\partial}{\partial
\vec{p}_i} \cdot \frac{1}{2} \Gamma_{ij}^2 \hat{r}_{ij}\hat{r}_{ij} \cdot
\vec{\cal L}_{ij} + \frac{\partial}{\partial u_i} \left[
\frac{1}{2m}(\vec{p}_j-\vec{p}_i) \cdot \frac{1}{2} \Gamma_{ij}^2
\hat{r}_{ij}\hat{r}_{ij} \cdot \vec{\cal L}_{ij} + \frac{1}{2}
\Lambda_{ij}^2 \left(\frac{\partial}{\partial u_i}-\frac{\partial}{\partial
u_j} \right) \right]\right\}  \nonumber \\
 & &  \label{19g}
\end{eqnarray}
where we have in addition defined the operator
\begin{equation}
\vec{\cal L}_{ij} \equiv \left(\frac{\partial}{\partial \vec{p}_i}
-\frac{\partial}{\partial \vec{p}_j} \right) +\frac{1}{2m}
(\vec{p}_j-\vec{p}_i) \left(\frac{\partial}{\partial
u_i}+\frac{\partial}{\partial 
u_j} \right) \label{19h}
\end{equation}
Note that the algorithm giving the aforementioned Fokker-Planck
equation is not unique. Effectively, the terms of order ${\cal O} (\delta t)$
in eqs.  (\ref{A1}), (\ref{A2}) and (\ref{A3}) can be replaced by their
averages over the random terms, giving a much simpler algorithm satisfying
the same Fokker-Planck 
equation, eq.  (\ref{19e}). Such a procedure was adopted in the first
version of the DPDE model\cite{ours}. However, replacing the drift terms
by their averages means that the energy is only conserved {\em in the mean}
but not
at {\em every time-step}. The implicit algorithm shown in eqs. (\ref{21a}),
(\ref{21b}) and (\ref{21c}) is thus a more compact form satisfying the
desired properties of the DPDE model described.

        The amplitudes of the random terms are derived by
imposing that the equilibrium distribution function given in eq. (\ref{Pe}) 
is a stationary solution of eq. (\ref{19e}), thus yielding the
fluctuation-dissipation theorems
\begin{eqnarray}
\Gamma_{ij}^2 &=& 2 k \, L_{ij}^{(p)} = 2 k \zeta_{ij} \,
\Theta_{ij}             \label{20} \\ 
\Lambda_{ij}^2 &=& 2 k \, L_{ij}^{(q)} = 2 k \theta_i \theta_j \, \lambda_{ij}
                        \label{20b} 
\end{eqnarray}
Eq. (\ref{19e}), in view of the definitions (\ref{19f}), (\ref{19g}) and
(\ref{19h}), together with the expressions for the amplitudes of the
random terms (\ref{20}) and (\ref{20b}), satisfies the expected
detailed balance condition\cite{JBA}. 

        The fluctuation-dissipation theorems given in eqs. (\ref{20}) and
(\ref{20b}) are the generalization of the results previously obtained in
ref.\cite{ours}, to the case of an arbitrary temperature-dependence of
the mesoscopic coefficients. We have to stress the fact that the
amplitude of the random force $\Gamma_{ij}$ depends only on particle
variables and that it is {\em independent of the macroscopic state} of the
system (the thermodynamic temperature, for instance). 
This point is a crucial difference between the DPDE model and
the previous isothermal DPD models, with no energy conservation. For the
isothermal DPD model, the 
amplitude of the random force is proportional to the thermodynamic
temperature of the system $T$ which must be previously specified. This
particular feature of the actual DPDE 
allows one to model temperature gradients in the systems as well as
heat flows. Furthermore, since the dynamics is based on particle
properties, DPDE can deal with systems either in contact with a heat
reservoir or with isolated systems. 

        Having set the
fluctuation-dissipation theorems that fix the amplitude of the random
terms, the algorithm describing the DPDE process
is completely specified. 

\section{Thermodynamics of the DPDE model}

\setcounter{equation}{0}

        The properties of the DPDE system defined so far are such that the
system evolves irreversibly towards a
final equilibrium state. If the system is in contact with a heat
reservoir, such an equilibrium state is characterized by the probability
distribution given in eq. 
(\ref{Pe}). Since this fact allows us to define a 
thermodynamics for the system of dissipative particles, we will devote this
section to the calculation of general results valid for any model.

        Let us introduce a partition function in a
classical sense, $Q(T,V,N)$, as 
\begin{eqnarray}
Q(T,V,N) &\equiv& \frac{1}{N! h^{3N}}\int \left(\prod_i d\vec{p}_i
d\vec{r}_i du_i \right)\; e^{-H(\{\vec{r}_i\},\{\vec{p}_i\})/kT}\prod_i
e^{s_i(u_i)/k-u_i/kT} \nonumber \\
 &\equiv& Q_H(T,V,N) \, Q_{int}(T,N)         \label{3b} 
\end{eqnarray}
where $Q_H$ refers to the partition function of the "hamiltonian" part of
the interaction and $Q_{int}$, to the partition function of the internal
degrees of freedom. The factor $1/h^{3N}$ has been introduced in analogy
with the partition function of a physical system.
The equivalent Free Energy for the system of
dissipative particles can thus be obtained as a sum
of two contributions 
\begin{equation}
{\cal F}(T,V,N) = -kT \ln Q_H(T,V,N) -kT \ln Q_{int}(T,N)     \label{FE}
\end{equation}
The former is related to the hamiltonian part of the dynamics and the latter,
related to the internal energy dynamics. $Q_H$, can be
obtained from the Hamiltonian of the system given in eq. (\ref{1}). One
gets 
\begin{eqnarray}
Q_H(T,V,N) &=& \frac{1}{N!h^{3N}} \int \left(\prod_i d\vec{p}_i
d\vec{r}_i \right)\; e^{-H(\{\vec{r}_i\},\{\vec{p}_i\})/kT} \nonumber \\
 &=& \frac{1}{\Lambda^{3N}N!} \int \prod_i
d\vec{r}_i 
\, e^{-\sum_i \left[\psi^{ext}(\vec{r}_i) + \sum_{j>i} \psi (r_{ij})
\right]/kT} \equiv \frac{1}{\Lambda^{3N}N!} Z_H(T,V,N) \nonumber \\
    \label{t1}
\end{eqnarray}
where $\Lambda \equiv h/\sqrt{2\pi m kT}$ is the de Broglie wavelength and
$Z_H(T,V,N)$ is the so-called {\em configurational integral}\cite{Han}.
The internal partition function can also be 
worked out. Effectively, one has
\begin{equation}
Q_{int}(T,N) = \int \prod_i du_i \, \prod_i e^{s_i(u_i)/k-u_i/kT} =
\left[\int du e^{s(u)/k-u/kT} \right]^N \equiv \left[Q_1(T) \right]^N
        \label{t10} 
\end{equation}
hence factorizing in one-particle partition functions, depending only on
the temperature. 

        Using these expressions, the free energy takes the form 
\begin{equation}
F(T,V,N) = kTN(\ln \rho +3 \ln \Lambda -1) -kT \ln \frac{Z_H(T,V,N)}{V^N}
        -kTN \ln Q_1(T) \label{t11}
\end{equation}
Note that $Q_{int}$ is a function of the temperature and the number of
particles, but is independent of the volume. Therefore, the pressure of
the system is completely determined by the Hamiltonian interactions
between particles, as one could have intuitively guessed. The internal
energy, however, contains contributions from all the terms.

        The explicit knowledge of the {\em equations
of state} of the system of DPD particles are very useful, since they can
be used to compare with the equations of state of a real system when
simulating
its behaviour and properties in a computer. In the rest of this section we
will derive general 
equations of state for the pressure and also for the internal energy of the
system, in terms of microscopic parameters of the model. The analysis
given here will follow that of ref.\cite{Han}, where the details of the
calculation can be found. Let us introduce the
one-particle and the two-particle distribution functions, according to the
relationships
\begin{equation}
\rho(\vec{r}) \equiv  \left \langle \sum_i \delta (\vec{r}-\vec{r}_i) \right
\rangle,         \label{t12}
\end{equation}
for the former, and
\begin{equation}
\rho^{(2)}(\vec{r},\vec{r} \prime) \equiv \left \langle \sum_{i,j\neq i} 
\delta (\vec{r}-\vec{r}_i) \delta (\vec{r}\prime-\vec{r}_j) \right \rangle
                \label{t13}
\end{equation}
for the latter. The averages refer to the equilibrium probability
distribution for the DPD system given in eq. (\ref{Pe}).
For a homogeneous and isotropic system, these expressions
can be further simplified. Effectively, the one-particle distribution
function is the mean density $\rho(\vec{r}) = N/V$ and the two-particle
distribution function reduces to 
\begin{equation}
\rho^{(2)}(|\vec{r}-\vec{r} \prime |) = \rho^2 g(|\vec{r}-\vec{r} \prime
|)      \label{t14}
\end{equation}
where this last equation is in fact a definition of the pair-distribution
function $g(r)$. 

        The pressure equation can be obtained from the so-called
{\em virial equation}
\begin{equation}
\frac{\beta P}{\rho} = 1+\frac{\beta}{3N} \left \langle \sum \vec{r}_i
\cdot \frac{\partial}{\partial \vec{r}_i} \sum_{j,k>j} \psi(r_{jk}) \right
\rangle         \label{t15}
\end{equation}
where $\beta \equiv 1/kT$. After some algebra, one arrives at the final
expression\cite{Han}
\begin{equation}
\frac{\beta P}{\rho} = 1 - \frac{2 \pi \beta \rho}{3}
\int_0^{\infty} dr \; r^3 \, \frac{\partial \psi (r)}{\partial r} g(r)
                        \label{t16}
\end{equation}
From this expression one can see that the actual model of DPDE particles
can exhibit the same phase behaviour as a fluid of pair interaction
potentials, thus {\em gas}, {\em liquid} and {\em solid} phases can be
modelled. The actual DPDE shares this property with the
older versions of the DPD model.

        The internal energy of the system can be
obtained by performing the equilibrium average of all the contributions to
this quantity
\begin{eqnarray}
U = \int d\vec{r} \left \langle \sum_i \frac{p_i^2}{2m} \delta
(\vec{r}-\vec{r}_i) \right \rangle &+& \int d\vec{r} d\vec{r}\prime \, \left
\langle \sum_{i,j>i} 
\psi(r_{ij}) \delta(\vec{r}-\vec{r}_i) \delta(\vec{r} \prime -\vec{r}_j) 
\right \rangle \nonumber \\
 & + & \int d\vec{r} \, \left \langle \sum_i u_i \, \delta (\vec{r}-\vec{r}_i)
\right \rangle          \label{t17}
\end{eqnarray}
Again, after some algebra, we obtain
\begin{equation}
U = \frac{3}{2} NkT +N\Psi + Ne_u       \label{t18}
\end{equation}
where the first term on the right hand side of this equation is the ideal
gas contribution. The second term is the interaction potential energy
contribution, and is given by
\begin{equation}
\Psi \equiv 2 \pi \frac{N}{V} \int dr \, r ^2 g(r)
\psi (r)  \label{t19}
\end{equation}
The third term is the contribution due to the internal energy of the
particles to the macroscopic internal energy of the system. It is
evaluated from the integral
\begin{equation}
e_u = \frac{\int_0^{\infty} du \, u \, e^{s(u)/k-u/kT}}{\int_0^{\infty} du
\, e^{s(u)/k-u/kT}}         \label{t20} 
\end{equation}
Note that the thermal capacity of the system is obtained by
differentiation, $\partial U/\partial T)_v$. This can be used to relate
the function $s(u)$, that has to be supplied to the model, with the
thermal behaviour of a given real system. These expressions cannot be
further developed without explicit calculation 
of the pair distribution function. However, they can shed some light on
the relevant aspects of the thermodynamic equilibrium in the DPDE system. 

        To end this section, let us introduce a non-equilibrium extension
of the free energy, given by the expression
\begin{equation}
{\cal F}[P] \equiv \int dX \, P(X,t) \left\{ f(X) +kT \ln P(X,t) \right\}
        \label{t22}
\end{equation}
which is defined as a functional of the actual probability distribution
$P$, and where $X$ is a point in phase space,
$(\{\vec{p}_i\},\{\vec{r}_i\},\{u_i\})$, and the function $f(X)$ is
defined as
\begin{equation}
f(X) = H(\{\vec{p}_i\},\{\vec{r}_i\}) + \sum_i \left(u_i-Ts_i(u_i) \right)
= -kT \ln P_e(X)
        \label{t23}
\end{equation}
Using this expression for the function $f(X)$, the eq. (\ref{t22}) can be
cast in a more convenient form
\begin{equation}
{\cal F}[P] = \int dX \, kT \, P(X,t) \ln \frac{P(X,t)}{P_e(X)} \label{t24}
\end{equation}
This functional ${\cal F}$ satisfies an $H$-theorem, $\partial {\cal
F}/\partial t \leq 0$, as follows from differentiation of eq. (\ref{t24})
with respect to time, with the use of the Fokker-Planck equation given in eq.
(\ref{19e}) together with the fluctuation-dissipation theorems (\ref{20}) and
(\ref{20b}).

\subsection{Analysis of a particular system}

        In the remainder of this section we will
analyze a particular model and show some simulation results. 

        Let us consider the case
in which the DPDE particles have no interaction potential
forces. The definition of the model requires a {\em particle equation of
state} relating the particle's entropy with the particle's internal energy,
$s(u_i)$. The simplest model is to assume that the particle temperature
is a linear function of the particle internal energy of the form $u_i =
\phi \, \theta_i$, where $\phi$ is a constant playing the role of a
particle's heat capacity. We will take $\phi$ to be larger than the Boltzmann
constant $k$\cite{gau}. Hence, the particle internal energy probability
distribution takes the form
\begin{equation}
P_e(u_i) \sim e^{s(u_i)/k-u_i/kT} = u^{\phi/k} e^{-u_i/kT}    \label{41}
\end{equation}
In the absence of an interparticle interaction potential the position
probability distribution is uniform, and the momentum
distribution is that of Maxwell-Boltzmann
\begin{equation}
P_e(\vec{p}_i) \sim e^{-\frac{p_i^2}{2mkT}}     \label{41b}
\end{equation}
The model is then
analogous to a general ideal gas and, hence, it 
corresponds to a
single gas phase, as seen from the resulting pressure equation
\begin{equation}
p = kT \rho,    \label{41c}
\end{equation}
in view of eq. (\ref{t16}). The second equation of state
determining the thermodynamic state of the system, follows from eq.
(\ref{t18}) and gives the variation of the macroscopic internal energy with
respect to the state variables. For our particular model, one gets
\begin{equation}
U = \frac{3}{2} NkT + NkT \left(\frac{\phi}{k} + 1\right) 
                \label{42} 
\end{equation}
In this equation, the first term is the ideal gas contribution to the
macroscopic internal energy, due to the translational degrees of freedom
of the particles. The second term, $\phi T$, is the internal
energy stored per particle. Note, however, the additonal $kT$ on the right
hand side of eq. (\ref{42}), that comes from the extra degree of freedom
due to the fluctuations in the internal energy of the
particles. The heat capacity can be calculated by differentiating eq.
(\ref{42}) with respect to the temperature, giving
\begin{equation}
C_v = Nk \left(\frac{\phi}{k} + \frac{5}{2} \right)     \label{42b}
\end{equation}
Let us point out that eq. (\ref{42b}) is a relationship between
macroscopically measurable magnitudes of any real system, $C_v$, with model
parameters $\phi$. 

        As far as the dynamic properties are concerned, the particle
friction and thermal conductivity are given by the expressions
\begin{eqnarray}
\zeta_{ij} & = & \zeta_0 \left( 1-\frac{r_{ij}}{r_{\zeta}} \right)^2 \;\;
\mbox{for} \;\; r_{ij} \leq r_{\zeta} \label{43} \\
\lambda_{ij} & = & \frac{L_0^{(q)}}{\theta_i \theta_j} \left(
1-\frac{r_{ij}}{r_{\lambda}} \right)^2 \;\; \mbox{for} \;\; r_{ij} \leq
r_{\lambda}     \label{44} 
\end{eqnarray}
where $\zeta_0$ and $L_0^{(q)}$ are constants giving the magnitude of the
mesoscopic friction and thermal conductivity, respectively, and
$r_{\zeta}$ and $r_{\lambda}$ are
the respective ranges of the dissipative interactions. The functions
$\zeta_{ij}$ and $\lambda_{ij}$ vanish if $r_{ij}>r_{\zeta}$ and
$r_{ij}>r_{\lambda}$, respectively. Although other
choices could be made for the spatial dependence of $\zeta_{ij}$ and
$\lambda_{ij}$, we have made here the usual choice\cite{Hoo}.

        The explicit derivation of the {\em macroscopic} transport
properties of 
the system will be treated elsewhere. Here, however, we give results that
can be easily calculated by considering that the transport of momenta and
energy is dominated by the dissipative interactions. Such a limit must be 
reached under conditions of high density system or by choosing large
particle friction 
and thermal conductivity parameters. By considering the particles as frozen
and analyzing the transport of mesoscopic heat through a hypothetical plane
dividing the system into two parts\cite{Doi}, one obtains for the
macroscopic thermal conductivity in these limiting conditions
\begin{equation}
\tilde{\lambda} \simeq \frac{\rho^2}{T^2} \frac{2 \pi}{3}
\int_0^{r_{\lambda}} dr \, r^4 \, L_0^{(q)} \left(
1-\frac{r_{ij}}{r_{\lambda}} \right)^2
g(r)  \label{35e}
\end{equation}
A similar calculation for the transport of momentum yields the shear
viscosity 
\begin{equation}
\eta \simeq \frac{2 \pi \rho^2}{15} \int_0^{r_{\zeta}} dr \, r^4 \,
\zeta_0 \left( 
1-\frac{r_{ij}}{r_{\zeta}} \right)^2 \, g(r)    \label{35f}
\end{equation}
which coincides with the result obtained in ref.\cite{Ern} when the
particle friction coefficient is considered as very large.
Performing the integrals given in eqs. (\ref{35e}) and
(\ref{35f}) 
by using the fact that in equilibrium and in the absence of pair
interaction potentials, $g(r) = 1$, we then obtain for the viscosity
\begin{equation}
\eta = \frac{2 \pi \rho^2}{1575} r_{\zeta}^5 \zeta_0     \label{46}
\end{equation}
For the macroscopic thermal conductivity, one gets
\begin{equation}
\tilde{\lambda} =  \frac{2 \pi}{315}\frac{\rho^2}{T^2}
L_0^{(q)} r_{\lambda}^5         \label{48}
\end{equation}
The analysis of the simulation results with steady state heat conduction
shows agreement with the functional dependence given in eq. (\ref{48}). A 
future publication will be devoted to a deeper analysis of the transport
properties of the DPDE model. 

        The velocity of sound is theoretically obtained by means of a
thermodynamic calculation
\begin{equation}
c^2 = \left. \frac{\partial P}{\partial \rho} \right)_{S} = \left.
\frac{C_p}{C_v}\frac{\partial P}{\partial \rho} \right)_{T} 
        \label{49} 
\end{equation}
For the case under discussion, one obtains for the speed of sound the
ideal gas result
\begin{equation}
c^2 = \frac{\phi+7k/2}{\phi+5k/2} \; \frac{kT}{m}  \label{50}
\end{equation}

        In order to carry out the simulations, we have introduced
dimensionless variables suitable for a proper interpretation of the
results. We have defined a temperature of reference $T_R$ to be used as
the scale of temperature. Thus, $T = T_R T^*$ and $\theta_i = T_R
\theta_i^*$, where the asterisk is used to denote a dimensionless
variable from now on. The momentum of the particles is made dimensionless
according with $\vec{p}_i =  \vec{p}_i^* \sqrt{2mkT_R}$, using the
characteristic value of the momentum in a system with no externally
imposed flow and thermal fluctuations: $\sqrt{\langle p_i^2 \rangle}$. The
penetration depth for the momentum defines a characteristic length scale
$l =\sqrt{2mkT_R} / \zeta_0$.
Thus, $\vec{r}_i = \vec{r}_i^* l $. The
characteristic scale of time is the relaxation time for the particle's
momentum, that is $t= t^* m/\zeta_0$. The
characteristic scale for the particle's internal energy is chosen to be
$\phi T_R$ so that $u_i = u_i^* \phi T_R$. When the algorithm is written
in terms of these dimensionless variables it can be seen that there are
only two independent dimensionless parameters in the model described in
this section, that is 
\begin{eqnarray}
B &\equiv& \frac{k}{\phi}       \label{49b} \\
C &\equiv& \frac{m L_0^{(q)}}{\zeta_0 \phi T_R^2} \label{50b}
\end{eqnarray}
$B$ measures the relative magnitude of the fluctuations in the particle's
internal energy with the characteristic particle energy. It is thus
convenient that $B$ be small. $C$ is the ratio between a
relaxation time for momentum decay $m/\zeta_0$, and the relaxation time
for the decay of a particle's internal energy fluctuations $\phi
T_R^2/\L_0^{(q)}$. Note that, to avoid negative values of the particle's
internal 
energy during the integration of the algorithm (eqs. (\ref{21a}),
(\ref{21b} and (\ref{21c})), in the
presence of externally imposed temperature gradients, one can
roughly estimate that 
\begin{equation}
C \left(\frac{4 \pi}{3} r_{\lambda}^{* 3} \right) \rho^*
\left(r_{\lambda}^* \nabla^* T^* \right) \delta t^* \ll 1  \label{51}
\end{equation}
where $\rho^*$ is the dimensionless particle number density and $\delta
t^*$ is the dimensionless integration time step. The fluctuating part of
the energy can also lead to negative energy values if the condition
\begin{equation}
\sqrt{BC \delta t^*} \left(\frac{4 \pi}{3} r_{\lambda}^{* 3} \right)
\rho^* \ll 1         \label{52}
\end{equation}
is not satisfied. 

        Two aspects of the DPDE model have been
investigated using simulations in three dimensions. In the first place,
the thermodynamic consistency has been checked by measuring the
equilibrium distributions for a system in contact with a heat reservoir.
Figs. 1 and 2 show the momentum and
particle's 
internal energy distributions of a system of $N=30000$ particles in contact
with two walls at a temperature $T^*=1.1$, while periodic boundary conditions
are chosen for the other two dimensions in space. The factor $B$ has been
set equal to $0.01$, $C=17$ and the cutoff lengths $r_{\lambda}^*$ and
$r_{\zeta}^*$ are both set equal to $1.24$. The size of the system is chosen
such that $\rho^*=1$, so that the dimensionless lateral size of the box is
$L^* = N^{1/3}$. The number 
of particles interacting simultaneously with a given particle is thus
determined by the size of $r_{\lambda}^*$ and $r_{\zeta}^*$. For our
values, we can roughly estimate that $8$ particles interact at one time
with a given particle. 

        The simulation has been initialised by a random distribution of
particles at rest and a dimensionless particle temperature $\theta_i^* = 1$.
At the initial stages of
the simulation, the system is cooled down to a temperature of about $0.98$
by the transformation of internal
energy into kinetic energy due to the action of the random forces, in a
time scale $t^*$ of about $1$. The difference in temperature between the
wall and the bulk provokes a heat flow from the walls to the particles 
(see Fig 3).
This process takes place in a much longer time scale, found of the order
of $100$, although this is clearly related to the overall size of the
system. In Fig. 3 we show the time evolution of the mean temperature of the
particles.  After equilibration, this mean temperature reaches the same
temperature as that of the walls.  In addition,
the three translational degrees of freedom are found to satisfy the
probability  
distribution given in eq. (\ref{41}) which, expressed in dimensionless
form reads
\begin{equation}
P_e(p_{i,\alpha}^*) = \frac{1}{\sqrt{\pi T^*}} e^{-p_{i,\alpha}^{* 2}/T^*}
                \label{53}
\end{equation}
where $p_{i,\alpha}^*$ stands for any of the three components, $\alpha$,
of the 
momentum of the i$^{th}$ particle, and $T^*$ is equal to the temperature of
the wall, $1.1$. In Fig.1,
simulation and theoretical results are shown together. We find excellent
agreement between simulated and theoretical distributions which, in
addition, are rather insensitive to the time step, provided it is small
($\delta t^* = 0.01$ in this simulation).  In equilibrium, the average 
internal energy
of the system is found to be $U^*=33828 \pm 1$. The same property 
calculated by 
eq.(\ref{42}) (written in dimensionless form) yields the value $U^*=33825$,
showing excellent agreement with the 
simulation value.

        The degree of freedom represented by the particle's internal
energy also satisfies the probability distribution given in eq.
(\ref{41b}). This distribution can also be cast in dimensionless form,
yielding 
\begin{equation}
P_e(u_i^*) = \frac{u_i^{* \, 1/B} e^{-u_i^*/BT^*}}{(B T^*)^{\frac{B}{B+1}}
\Gamma \left(\frac{B}{B+1} \right)}     \label{54}
\end{equation}
where here $\Gamma$ stands for the Euler Gamma function, and $T^*$
is the temperature of the walls. Fig. 2, shows the theoretical predictions
and simulation results for this probability distribution. Once more, the
agreement is excellent. This result is non trivial in view of the fact
that the functional form of the probability distribution for the energy is
arbitrary, and depends on the choice made for the particle equation of
state, $u_i = \phi \theta_i$ in our case. We have found, however, that
this probability distribution is much more sensitive to the time-step
($\delta t^* = 0.0001$ in the simulation shown in Fig. 2). For a time step
$\delta t^* = 0.01$, the deviations are less than 1\%. 

        Finally, the pressure in
the simulation can be obtained from the average force that the particles
exert on 
the walls.  For the previously mentioned system we have obtained the pressure
from the simulation and found a value of $P^*=1.137\pm 0.005$.  The theoretical
value given from the equation of state eq. (\ref{41c}) is $P^*=1.1$.  Again
we find a good agreement between simulation and theory.

        Simulations of
closed systems (with periodic boundary conditions in the three dimensions)
have also been performed (this is possible because the algorithm depends 
on particle properties only). The DPDE system tends in this case towards a
final thermodynamic equilibrium satisfying the aforementioned probability
distributions for the one-particle variables. In this case, however, the
probability distribution of the complete system is not separable as it was
in the previous case. The final temperature attained for the system is in
excellent agreement with that predicted from eq. (\ref{42}), provided that
the total energy of the system is known. 

        The second aspect analysed is the ability of the model to
represent non-equilibrium features of fluids under temperature gradients,
an unattainable problem for the older DPD algorithms. We have performed
simulations of a DPDE system with $N=100000$ particles in a cubic box of
lateral size $N^{1/3}$, with four walls and periodic boundary conditions
in the remaining direction.
The walls were considered to be a two-dimensional surface which exert a
repulsive force in the orthogonal direction on the particles.  The range
of this force is of the order of the range of the particle-particle
interaction.  As far as the wall-particle dissipative interactions are
concerned, we have considered as being analogous to the particle-particle
interactions.  Hence when a particle is within the interaction range of a 
wall, it exchanges heat and exerts a force orthogonal to the wall as if the
latter were a particle of
infinite mass and heat capacity.  Therefore, the walls act as heat 
reservoirs at a given specified temperature, and non-stick boundary
conditions apply.

In the simulations, the two walls in the $x$-axis were kept at fixed
temperatures of $T_h^* = 2.8$ and $T_c^*=0.8$, while the other two walls
were adiabatic and were placed in the $z$-axis, along which a gravity field
was imposed on the particles. Qualitatively, the gravity was weak enough
not to cause an excesive density gradient along the $z$-axis, and the
particle's thermal condutivity was chosen to be very small to emphasize
the convective effects. Initially at rest, the system spontaneously
evolved towards a stationary convection roll as seen in Fig. 4, where the
arrows stand for the fluid velocity field. The velocity field has been
calculated by dividing the space into boxes and averaging the velocity of
the particles inside the box at a given instance of time. The fields
obtained in this way show a rather marked variability due to the inherent
stochastic nature of the algorithm and the limited number of particles
inside a given box at a given time. The results shown in Fig. 4 are thus
the result of averaging in time as well as in the $y$-direction, to have
better statistics. In addition, the temperature profile has been obtained.
We see in Fig. 5 that a non-linear temperature profile, due to the
compressibility of the model, exists and is slightly tilted due to the
convective flow. These results are in qualitative agreement with
convective rolls observed in boxes and in numerical solutions of the
Navier-Stokes equations for incompressible systems under the conditions
described here. The estimated Rayleigh number is of the order of $1000$,
by comparison with the numerical solution of the Navier-Stokes equations.
In Fig. 6 we show the density profile given by the model, in which one can
appreciate the combined effects of the temperature gradient and gravity
field, merging into a diagonal density gradient, due to the
compressibility of the system. It is interesting to note how this density
profile affects the convection pattern in the system (Fig. 4), causing a
displacement of the vortex centre towards the denser region, and also
inducing a faster motion in the upper layers, as compared with the bottom
of the box.

\section{Conclusions}

\setcounter{equation}{0}

        Various aspects of the {\em Dissipative Particle Dynamics with
Energy Conservation} algorithm, already introduced in
ref.\cite{ours}, have been treated here in depth. In particular, emphasis
has been placed on two major points. Firstly, the original DPDE
algorithm has been extended in two ways: to incorporate arbitrary
temperature-dependencies in the transport coefficients, and to assure that
the algorithm conserves the energy at every time-step rather than in the
mean, as in our previous algorithm\cite{ours}. Secondly, in this paper we
have studied the thermodynamic properties modelled by the DPDE
algorithm such as the free energy and equations of state, and shown
agreement between the simulated and theoretically predicted probability
distributions. To demonstrate the ability of the DPDE algorithm in simulating
the hydrodynamic behaviour of fluids under non-equilibrium conditions,
simulations have been carried out for a system with a temperature gradient
orthogonal to a gravity field. The results show the expected convective
pattern for the velocity field, as well as the corresponding tilted
temperature profile. These points stress the inherent features of
the new DPDE algorithm as compared with the isothermal versions.

        With respect to the first major point, and with a view to the use
of the DPD methodology for 
the simulation of the dynamics of real fluids, it was necessary to
incorporate arbitrary dependencies of the transport
coefficients in the temperature. We have constructed an algorithm where
the mesoscopic friction and heat conduction coefficients, $\zeta_{ij}$ and
$\lambda_{ij}$, respectively, can depend on the particle's temperature.
One can compare eq. (\ref{43}) with  
eq. (\ref{35f}), to see that both coefficients are independent of the
temperature, neither the particle's nor the thermodynamic temperatures. 
Contrarily, the functional form of eq. (\ref{44}) gives rise to the
$1/T^2$ dependence of the thermal conductivity coefficient. Of course, since
eqs. (\ref{35e}) and (\ref{35f}) are approximate, there must exist additional
temperature dependencies that may be due either to $g(r)$ or to the
kinetic transport of momentum and energy. Note that, the latter has been
neglected in our derivation of the transport coefficients. 

        Another aspect worth mentioning in this context is our derivation
of an implicit algorithm. This derivation has two important 
properties.  On the one hand, the algorithm can be directly obtained from the
Langevin equations used to formulate the problem of the DPD
particle dynamics.  We have simply introduced an interpretation rule of
the random terms, when 
integrating the equations of motion in a $\delta t$, which is different from
the usual It\^o-Stratonovich interpretations. On the other hand, the resulting
algorithm, with the proper fluctuation-dissipation theorems,
straightforwardly satisfies detailed balance which is required for 
the model to behave in a thermodynamically consistent manner.
Furthermore, since this algorithm has been directly obtained from the
Langevin equations (\ref{4}), (\ref{5}), and (\ref{10}), it satisfies
energy conservation at every time step, as do the Langevin equations
themselves.

        We have also indicated that a family of different algorithms can
lead to 
the same Fokker-Planck equation. Since the process is Gaussian, it is
completely determined by the 
first and second moments of the probability distributions for the random
variables. Thus, the 
defining trends of this family of algorithms is that these first and
second moments be the same for all of them, up to the order of validity of
the algorithm itself, ${\cal O}(\delta t)$. As an example, let us average
eq. (\ref{A2}) with respect to the random number $\Omega_{ij}^{(p)}$, giving
\begin{equation}
\langle \delta \vec{p}_i \rangle = \vec{F}_{i}^{ext} + 
\sum_{j \neq i} \left[\vec{F}_{ij}^C + \left(\frac{\zeta_{ij}}{m} +
\frac{1}{2m} \Gamma_{ij} \left(\frac{\partial}{\partial u_i} +
\frac{\partial}{\partial u_j} \right) \Gamma_{ij}
\right)\, 
(\vec{p}_j-\vec{p}_i) \cdot \hat{r}_{ij} \hat{r}_{ij} \right]   \label{53b}
\end{equation}
The same result is obtained from our previous algorithm proposed in
eq. (16) of ref.\cite{ours}. This demonstrates that indeed the firsts
moments of the variation of the momentum for both algorithms are
identical. Of course, the same procedure can be repeated for all the first
and second moments of all the increments of the variables in a time-step.
Thus, despite the fact that the macroscopic
properties of both algorithms are the same, it should be emphasised
that the present algorithm preserves energy conservation at every
time-step while all the other possible algorithms only do so in the mean.
This can be 
verified by summing all the contributions to the energy for a pair of
particles in a time-step. This point is of
crucial importance when analysing the dynamic behaviour of 
the system, in particular when the exact conservation of the energy is
checked during the simulations.

        Regarding the second major point, we have emphasized in this
article the existing 
tight relation between the overall properties of the DPDE model developed
here and the macroscopic 
properties of real systems. This tight relationship will allow DPDE models to
simulate the 
dynamic behaviour of real fluids, with respect to the equilibrium and
transport properties of both momentum and energy. When the energy
conservation is 
introduced, the existence of a {\em Thermodynamic behaviour} of the DPDE
system becomes apparent. We have seen that the hypothesis leading to the
present algorithm can reproduce not only the Maxwell-Boltzmann probability
distribution for the momenta, but also the arbitrary probability
distribution for the particle's internal energy. Furthermore, we have
introduced 
a free energy from a partition function related to the system probability
distribution.  In the calculation of this partition function, the 
particle's internal energy $u_i$ has been considered as an additional
microscopic degree of 
freedom. From this formulation, we can obtain equations of state in
terms of the parameters of the model. In particular, we have obtained a
relationship between the internal energy of the whole system and its
macroscopic temperature. This equation of state is useful since it allows one
to relate 
the total energy introduced in a closed DPDE system with the final equilibrium
temperature. The agreement between the predicted temperature and the
simulated one is excellent. The pressure equation for this system has also
been obtained, showing once more excellent agreement between the simulated and
theoretical predictions.  The pressure in a DPD system in general
is constrained by the particle number density, which is much smaller than that
of the corresponding real system being modelled. This is due to the
implicit coarse-graining in the model, in which a packet of
physical molecules is represented by a single DPD particle.
Clearly, this is an area that needs to be further studied.  However, for
low Mach number flows, the DPDE fluid can be regarded as effectively
incompressible, 
and thus a Boussinesq-like approximation is of use. Under these
conditions, the thermal dilatation coefficient is the one that has to be
properly modelled. In the particular system analyzed here, for instance, this
coefficient, $\alpha$, is that of an ideal gas
\begin{equation}
\alpha \equiv \frac{1}{V} \left. \frac{\partial V}{\partial T} \right)_{p} =
\frac{1}{T}             \label{co1}
\end{equation}
With a proper choice of the interaction potential this coefficient could
be tuned according to the kind of fluid to be modelled. Note the excellent
qualitative agreement found between the non-equilibrium simulations of a
convective flow presented here and real fluid flows under equivalent
conditions. These simulations can be theoretically analysed in light of the
Boussinesq approximation.
       Finally, we have also identified a
non-equilibrium free energy which acts as a Lyapunov functional for the
dynamics of the system. Therefore, as seen in the simulations the
DPDE system has an inherent irreversible tendency towards a thermal
equilibrium when no external forcing exists. Moreover, other quantities
have been theoretically derived such as the 
viscosity, the thermal conductivity, and the speed of sound. An in
depth analysis of the derivation of these quantities and the consequences
that can be drawn from its dynamic behaviour lie beyond the scope 
of the present work.

        Therefore, the inclusion of the internal energy in the DPD
algorithm reinforces the internal consistency of the model and opens
interesting perspectives aimed at the use of these models 
in the simulation of the macroscopic-mesoscopic behaviour of simple and
complex fluids. 

\section*{Acknowledgements}

The authors wish to acknowledge the support of the grant PB96-1025, from the
Direcci\'on General de Ciencia y Tecnolog\'{\i}a
of the Spanish Government. JBA wishes to also
thank I. Pagonabarraga for stimulating discussions.

\newpage

\section*{Figure Captions}

\noindent Fig. 1:  Equilibrium momentum probability distribution
expressed in reduced variables, under the conditions ($T^*=1.1$, $N=30000$,
$C=17$, $B=0.01$,$r_{\lambda}=r_{\zeta}=1.24$). The solid line stands for
the theoretical prediction given in 
eq. (\ref{53}) and the dots are simulation results. The time-step is
$\delta t^* = 0.01$.

\noindent Fig. 2: Equilibrium particle's internal energy distribution
expressed in reduced variables, under the conditions
($T^*=1.1$, $N=30000$, $C=17$, $B=0.01$,$r_{\lambda}=r_{\zeta}=1.24$). The
solid line is a plot of the theoretical distribution given in eq.
(\ref{54}). The time-step used in this simulation is $\delta t^*=0.0001$.

\noindent Fig. 3: Equilibration of the temperature starting from a random
configuration under the same conditions as given in Fig. 1.

\noindent Fig. 4: Velocity field for a system under a temperature gradient
orthogonal to a gravity field. The wall temperatures were chosen to be
$T_h^*=2.8$, on the left hand side, and $T_c^* = 0.8$ on the right. The
gravity field was $g^* = 0.01$, directed downwards. The system
contains $N=100000$ particles in a cubic box of side $N^{1/3}$, with
adiabatic walls in the top and the bottom sides of the cube and periodic
boundary conditions in the remaining direction. The additional parameters
were chosen $B=0.1$, $C=0.001$, $r_{\zeta}=r_{\lambda}=2$, with a
time-step $\delta t^* = 0.01$.

\noindent Fig. 5: Isothermal lines for the system described in Fig. 4. The
numbers stand for the thermodynamic temperature $T^*$ for a given line.

\noindent Fig. 6: Constant density lines for the system described in Fig.
3. The numbers represent the values of the density $\rho^*$ for a given
line. The distortion near the walls are caused by repulsive potentials
exerted by the walls used to confine the system.


\begin{thebibliography}{99}
\bibitem{JB4} The local equilibrium is the lowest order term in an
expansion of the Fokker-Plank equations essentially in a parameter scaling
as 1/friction (See, for instance, ref.\cite{vK} and references
there in). This expansion method applied to the calculation of transport
equations in a particle suspension can be found in refs. \cite{JB2,JB3}.
\bibitem{gau} The requirement that the heat capacity of the particle must
be larger than $k$ is due to the fact that the mesoscopic random heat flow
is modelled as a continuum process subjet to Gaussian statistics. Thus,
the probability that a given particle, $i$ say,  at a given instant of
time randomly loses more energy than the actual value $u_i$ is
exponentially reduced if $\phi > k$. This is due to the fact that the
characteristic particle 
internal energy is of the order of $\phi T$ while fluctuations in the same
variable are of the order of $kT$. Therefore, in general, the model is
only able to 
deal with "classical" or "high temperature" situations, where the DPD
particles have an average internal energy substantially larger than the
thermal energy $kT$.
\bibitem{DB} The relation between the equations of motion for a given system
and the detailed balance condition can be found, for instance, in \S V.6 of
reference\cite{vK}.
\end{thebibliography}
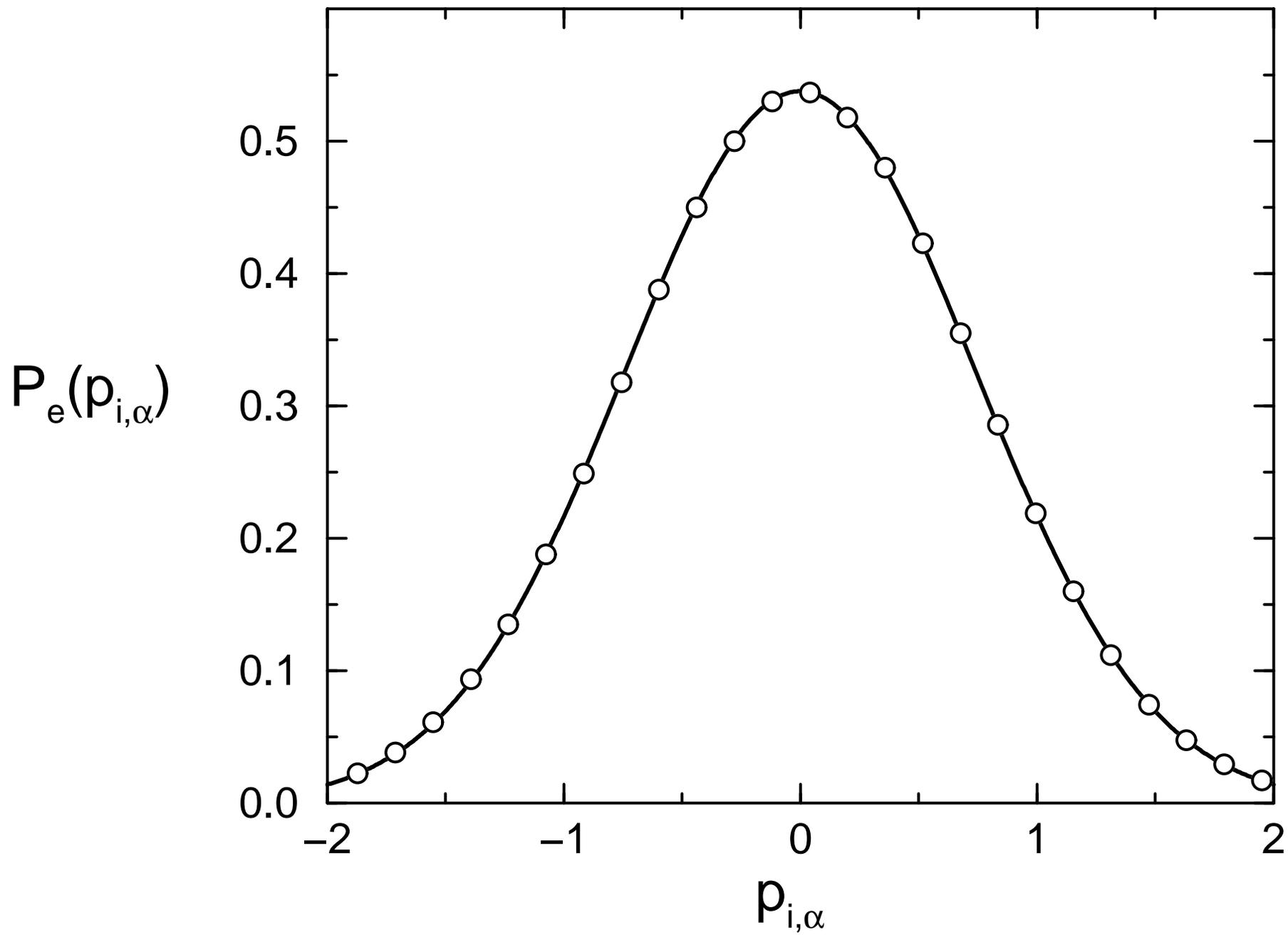
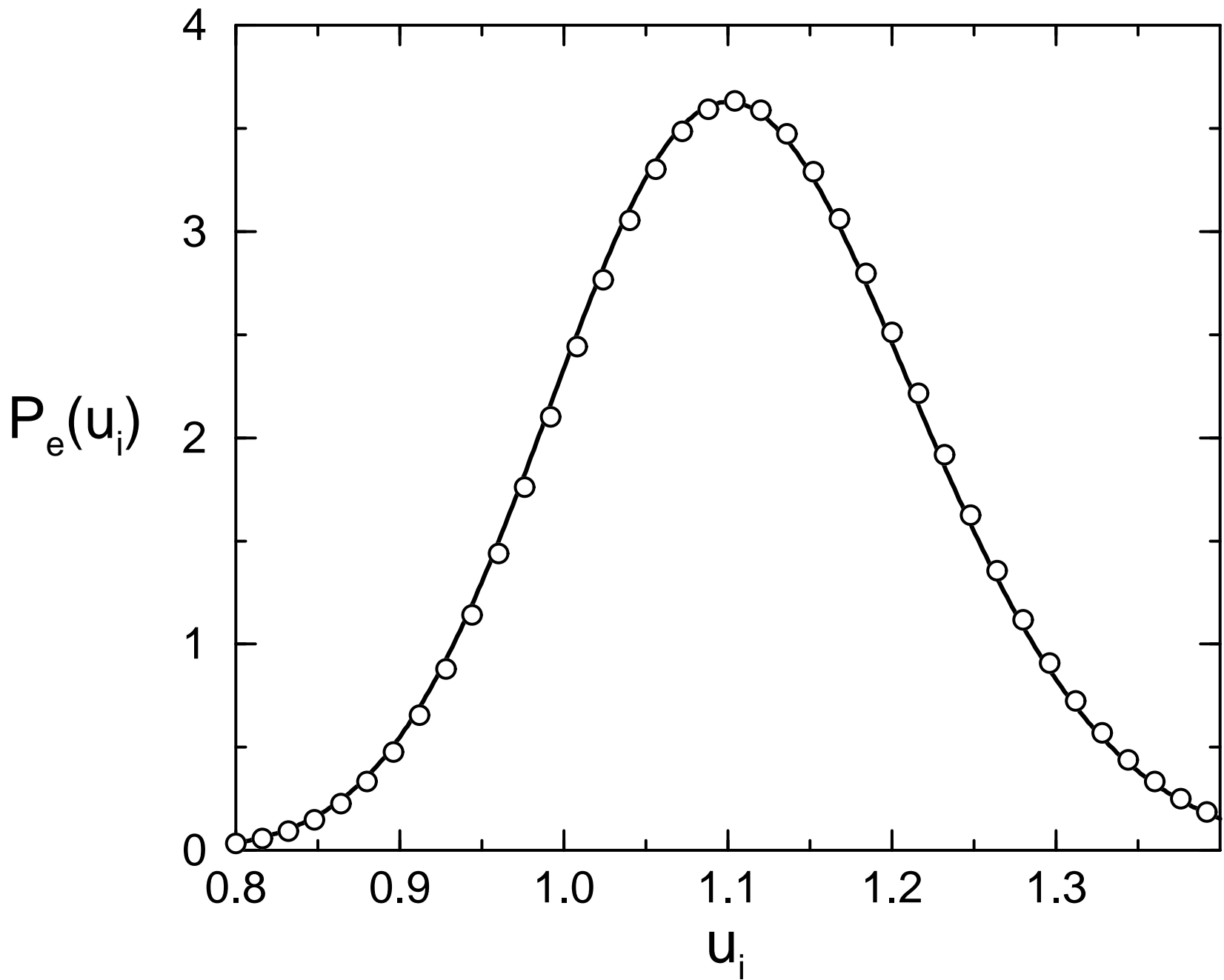
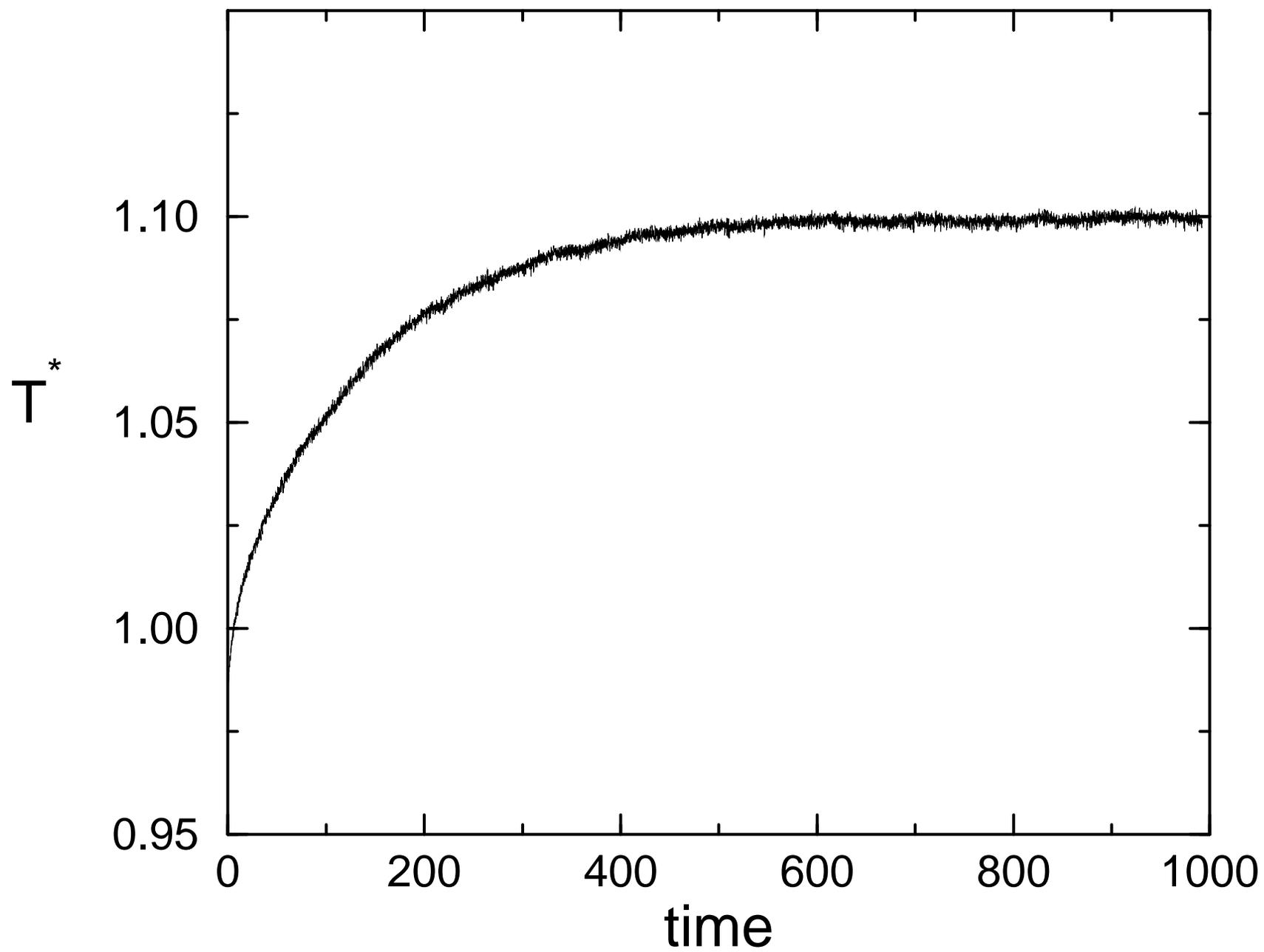
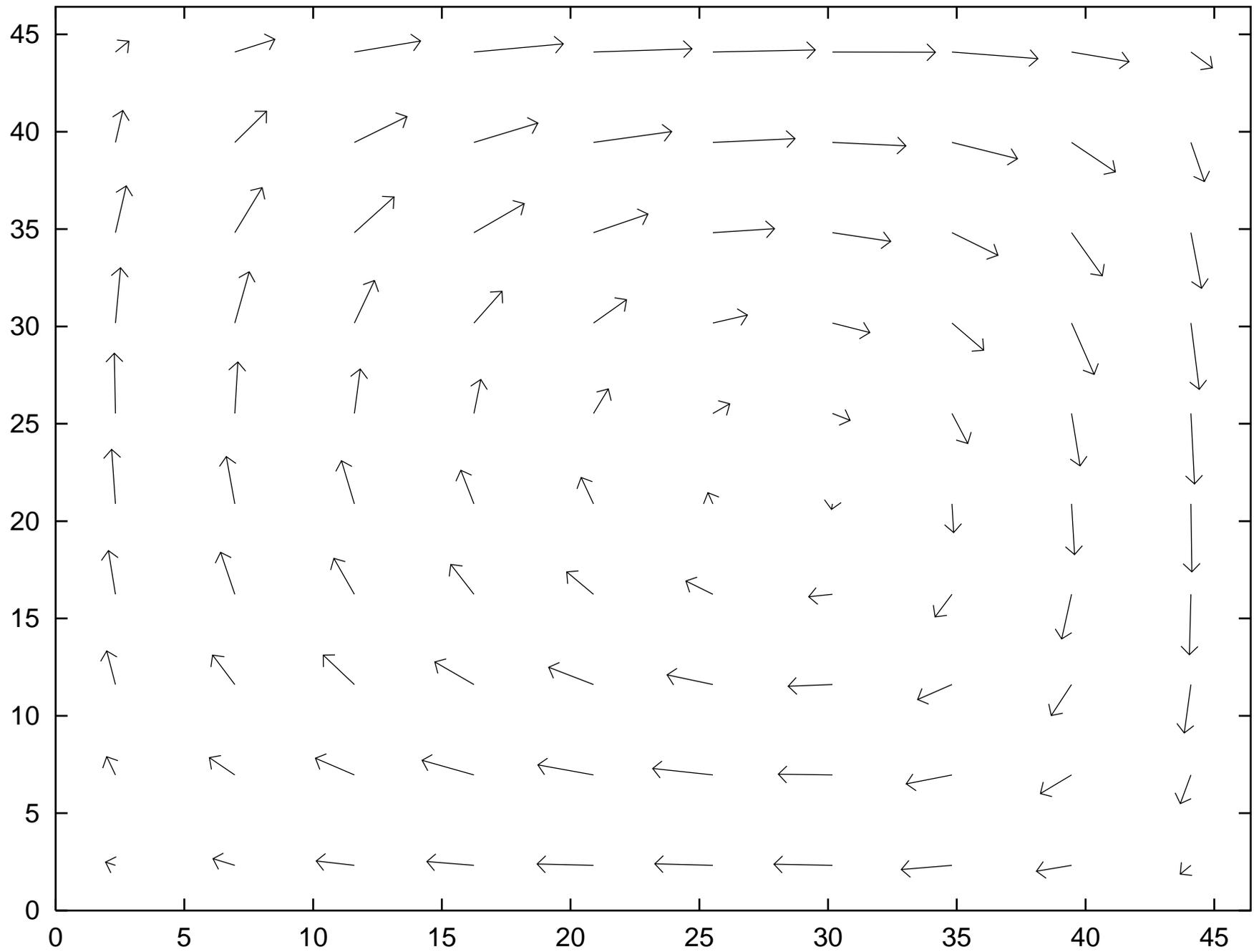
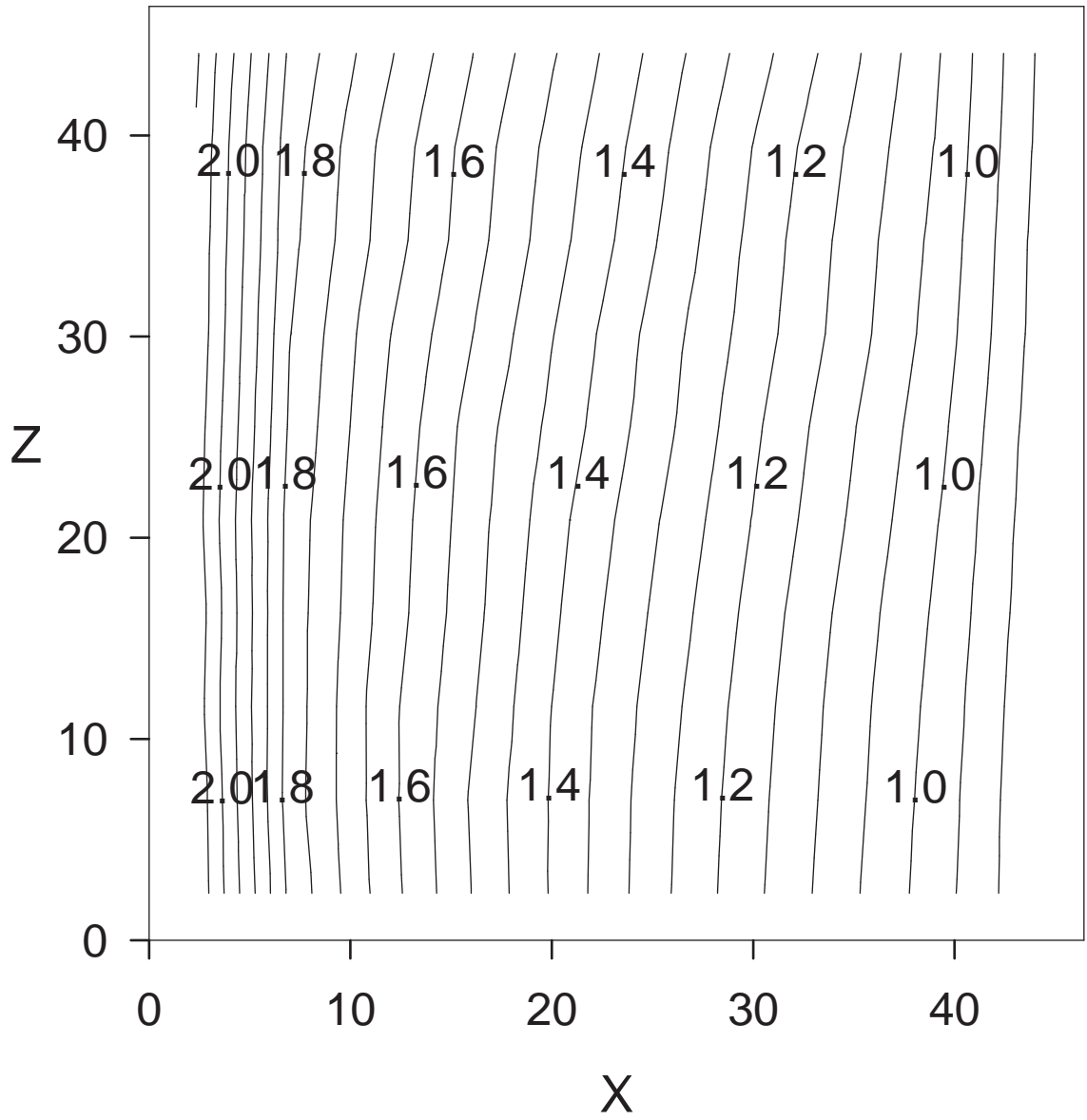
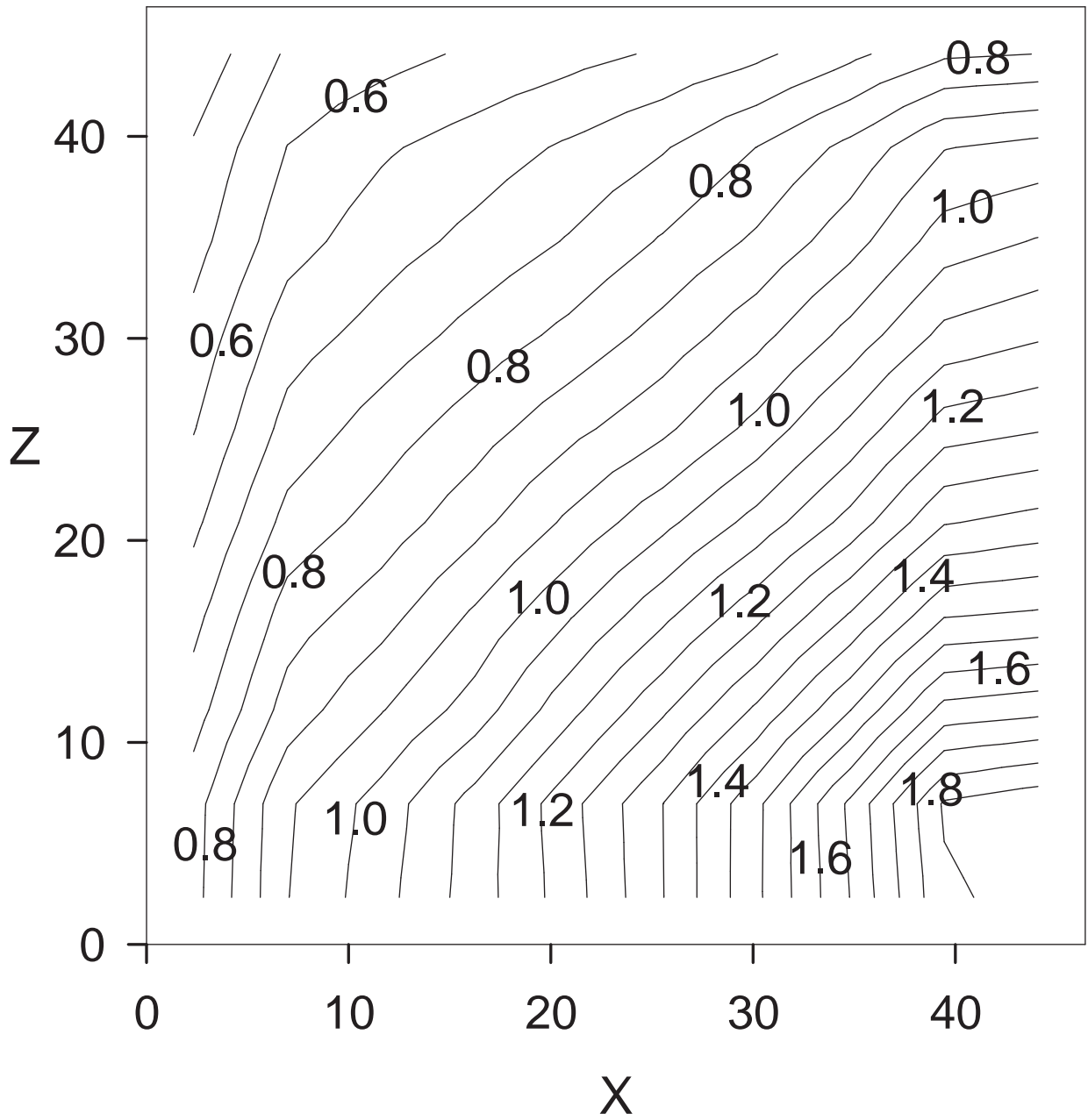
\end{document}